\documentclass[
reprint,
superscriptaddress,
longbibliography,
showkeys,
 amsmath,amssymb,
 aps,
onecolumn
]{revtex4-2}

\usepackage{graphicx}
\usepackage[caption=false]{subfig} 
\usepackage{amsmath}
\usepackage{lipsum}
\usepackage{amssymb}
\usepackage{mathtools}
\usepackage{ bbold }
\usepackage{physics}
\usepackage{float}
\usepackage{algorithm}
\usepackage{algpseudocode} 
     
\usepackage{dcolumn}
\usepackage{bm}
\usepackage{hyperref}

\newcommand{\ee}{\end{equation}} 
\newcommand{\be}{\begin{equation}}

\usepackage[mathscr]{euscript}  
\usepackage{xcolor}  
\usepackage{tcolorbox}
\usepackage{comment}
\usepackage{float}

\definecolor{myorange}{RGB}{220,90,30}

\newcommand{\newtext}[1]{{\color{black}{#1}}}
\newcommand{\deletetext}[1]{}

\makeatletter
\newsavebox{\@brx}
\newcommand{\llangle}[1][]{\savebox{\@brx}{\(\m@th{#1\langle}\)}%
  \mathopen{\copy\@brx\kern-0.5\wd\@brx\usebox{\@brx}}}
\newcommand{\rrangle}[1][]{\savebox{\@brx}{\(\m@th{#1\rangle}\)}%
  \mathclose{\copy\@brx\kern-0.5\wd\@brx\usebox{\@brx}}}
\makeatother

\begin{document}

\title{Rare events of small-noise Doob conditioned processes}

\author{Iago N.\ Mamede}
\thanks{\textit{iagomamede@usp.br}}
\affiliation{Instituto de F\'{i}sica da Universidade de S\~{a}o Paulo, 05314-970 S\~{a}o Paulo, Brazil}
\affiliation{UHasselt, Faculty of Sciences, Theory Lab, Agoralaan, 3590 Diepenbeek, Belgium}

\author{Francesco Coghi}
\thanks{\textit{francesco.coghi@leicester.ac.uk}}
\affiliation{School of Physics and Astronomy, University of Nottingham, Nottingham, NG7 2RD, United Kingdom}
\affiliation{Centre for the Mathematics and Theoretical Physics of Quantum Non-Equilibrium Systems, University of Nottingham, Nottingham, NG7 2RD, United Kingdom}
\affiliation{School of Computing and Mathematical Sciences, University of Leicester, LE1 7RH, Leicester, United Kingdom}
\affiliation{The Alan Turing Institute, NW1 2DB, London, United Kingdom}

\begin{abstract}
Doob fixed-time conditioning enables the sampling of rare trajectories of Markov processes by modifying the drift so that reaching a prescribed target at a given time is guaranteed. We study the statistics of this conditioned path ensemble through the moment generating function in the weak-noise large deviation regime. Since the Doob drift is rarely available in closed form, we reinterpret the conditioned ensemble as the original process post-selected on the terminal constraint, thereby avoiding explicit construction of the Doob transform. This viewpoint then yields an optimal-control representation for the leading exponential contribution to the generating function, expressed as a variational principle with terminal boundary conditions set by the Doob end-point constraint. We illustrate the framework with two analytical examples and with an application to heat dissipation \newtext{under Doob-controlled dynamics in}\deletetext{of} a minimal model of biomolecular folding.
\end{abstract}

\maketitle

\section{Introduction}

Having access to specific behaviours and dynamical regimes of processes influenced by noise is a central objective in the study of stochastic dynamics. This objective becomes increasingly important when the specific behaviour or event we are interested in is far from typical as, although rarely seen, atypical events \deletetext{may lead to drastic changes in the dynamics that} generate long-time correlations potentially leading to cascading effects in the future~\cite{Albeverio2006,Scheffer2009}. However important they are, ensembles of rare trajectories are in general hard to access\deletetext{, both experimentally and computationally}. Experiments are typically confined to typical behaviour and typical fluctuations of noisy systems, although carefully designed small-system experiments in stochastic thermodynamics have made it possible to access rare fluctuations at the trajectory level~\cite{Wang2002,Ciliberto2017}. Similarly, direct numerical simulations in silico\deletetext{, such as na\"{i}ve Monte Carlo sampling,} are naturally confined to typical trajectories. Accessing rare events by brute force requires a computational effort that grows inversely with their probability, and therefore exponentially in the large-deviation regime. This difficulty has motivated rare-event theory to branch into several directions, leading to efficient numerical methods to sample rare trajectories~\cite{Bucklew2004,Giardina2011,Hartmann2013,Bouchet2019,Grafke2019}.

Among all possible ensembles of rare trajectories one might be interested in, fixed-time conditioning a noisy system at a specific position in the state space has a prominent role. If the original system is Markov, then this setup can be seen as an instance of Doob's original conditioning problem, which dealt with exiting a certain domain via a subset of its boundary at a random (first-passage) time~\cite{Doob1957,Doob1984}. The outstanding result of Doob is that it is possible to derive a new conditioned process via the so-called Doob's $h$-transform which introduces a new drift---in terms of the so-called Doob's $h$-function---but keeps the same noise amplitude as well as preserves Markovianity. Doob's framework is the mathematical framework of stochastic/Langevin bridges~\cite{Orland2011,Majumdar2015,Aguilar2022} as well as Schr\"{o}dinger bridges~\cite{Lonard2013} where the process is conditioned on reaching a target distribution rather than a target point. Relevant \newtext{physical} examples \newtext{arise in systems displaying noise-induced transitions between metastable states, where one is interested in trajectories realising a prescribed transition within a finite time. These} include\deletetext{studying the finite-time dynamics of noisy systems moving between multiple basins of attraction such as} icehouse-greenhouse transitions in climate systems~\cite{Lucarini2019}\newtext{,}\deletetext{ and} \deletetext{open-to-native folding in proteins~\cite{Onuchic1997EnergyLandscape}} \newtext{conformational transitions in proteins~\cite{Onuchic1997EnergyLandscape}, and other general reactive paths~\cite{Crou2013}.}

\deletetext{Once the ensemble of these rare trajectories is identified and reconstructed via Doob's theory, one has access to its statistics, such as the typical conditioned dynamics as well as to the fluctuations of these conditioned processes and observables associated to them~\cite{Touchette2009,Jack2010,Garrahan2016,Touchette2018}. Fluctuations are in particular important because they help clarify whether the rare event is realised via a unique mechanism---then the typical rare-event dynamics is well representative---or multiple mechanisms/metastable dynamics, leading potentially to coexistence of dynamical phases and biases in sampling the conditioned dynamics. All this information is provided by the moment generating function of the Doob conditioned process, which should, in principle, be accessible within the weak-noise large deviation framework~\cite{Freidlin1984}. However, the Doob's $h$-function drift entering the conditioned dynamics is rarely available analytically---it is the solution of the backward Kolmogorov equation of the original process~\cite{Chetrite2015,Majumdar2015}---causing technical difficulties in extracting information from the moment generating function of the conditioned process.}
\newtext{Once this rare-event ensemble is identified through Doob's theory, one can ask not only what a typical conditioned trajectory looks like, but also how physical observables fluctuate across the ensemble of trajectories realising the transition. Such observables may include \newtext{activity and }\deletetext{heat, work, entropy flow, activity, or} other time-integrated quantities~\cite{Touchette2009,Jack2010,Garrahan2016,Touchette2018}\newtext{, as well as thermodynamic observables such as heat or work when the Doob drift is interpreted as an external control}. Their fluctuations resolve structure that is invisible at the level of the typical conditioned dynamics alone: for instance, whether the prescribed transition is realised through a single dominant mechanism or through several competing dynamical pathways or metastable regimes. In this sense, the problem considered here is a large-deviation problem \emph{within} a rare-event ensemble: the Doob conditioning selects the transition of interest, while the fluctuations of a path observable characterise the different ways in which that transition can occur.

The weak-noise regime is particularly natural for this problem. When noise is small, transitions between metastable states are themselves rare and conditioned trajectories concentrate around optimal paths, or instantons, making the Freidlin--Wentzell large-deviation framework applicable~\cite{Freidlin1984}. The corresponding moment generating function of an observable of the Doob-conditioned process should therefore provide access to fluctuations around such dominant transition mechanisms. In practice, however, the Doob $h$-function entering the conditioned drift is rarely available analytically---it is determined by the backward Kolmogorov equation of the original process~\cite{Chetrite2015,Majumdar2015}---which makes a direct weak-noise analysis of the conditioned process technically difficult.}

Yet, the ensemble of trajectories generated by Doob's conditioned dynamics coincides with that of the original process post-selected on the constraint~\cite{Doob1957,Majumdar2015,Corstanje2023,Coghi2025}. Recasting the problem as post-selection lets us sidestep the technical difficulties of working directly with Doob's $h$-function and yields a convenient rewriting of the moment generating function of the conditioned process. This reformulation opens the door to further analytical progress. The moment generating function satisfies a generalised Feynman--Kac equation (for details on Feynman--Kac theory see~\cite{Kac1949,Fitzsimmons1999}), and its leading exponential behaviour---which we call the \emph{action}---is precisely the object from which we extract the desired generating function. We obtain it from a Hamilton--Jacobi equation supplemented by a new boundary term, a space-time constraint that plays the role of the Doob conditioning. 

Beyond simple 1D systems, this approach must be handled numerically. For complicated systems it also offers numerical advantages. As discussed in~\cite{Vanden-Eijnden2012}, rather than studying the characteristics of the Hamilton--Jacobi equation, one switches to the Lagrangian formulation and writes the action in its optimal\newtext{-}control representation, which is accessible through iterative numerical optimisation methods~\cite{E2004,Grafke2019}.

We also remark that our approach \newtext{fits within the broader programme of describing rare trajectory ensembles through effective dynamics and variational principles, and} is close in spirit to the weak-noise \deletetext{variational} analysis \deletetext{of large deviations conditioned} on large deviations developed in~\cite{Derrida2019}. The difference is that their conditioning is imposed on long-time empirical observables, whereas here the conditioning is a finite-time Doob bridge constraint. This leads to a Hamilton--Jacobi problem with a terminal boundary term representing the endpoint conditioning\deletetext{, and}\newtext{. The framework} allows us to study fluctuations of observables\deletetext{, such as heat, inside the} \newtext{within a prescribed} \deletetext{fixed}\newtext{finite}-time conditioned ensemble \newtext{and, when the Doob drift is interpreted as an external control, thermodynamic observables such as the heat dissipated by the corresponding controlled dynamics}.

The paper is structured as follows: in Sec.\ \ref{sec:Doob} we introduce Doob fixed-time conditioned processes and, for simplicity, only consider additive noise (in Appendix \ref{app:multi} we show the derivation of the generalised Feynman--Kac equation for multiplicative noise too) and discuss two well-known analytical examples. In Sec.\ \ref{sec:Weak} we introduce the time-additive observables of the conditioned processes whose statistics we are after and build the weak-noise large deviation framework for the moment generating function rephrased with the post-selection point of view. We introduce the action, derive the Hamilton--Jacobi and, eventually, the optimal control representation of the action. In Sec.\ \ref{sec:ex}, we discuss two analytical bridges and a minimal model of biomolecular folding, in which the system transitions from the unfolded state to the folded one via saddle-point crossing. As an observable, we \deletetext{will }specifically focus on the heat dissipated \newtext{under the Doob-controlled dynamics that realises the folding transition}\deletetext{ by the biomolecule as it folds as a function of time}. Finally, we conclude in Sec.\ \ref{sec:conc} with a summary and outlook for future research.

\section{Doob conditioned processes}
\label{sec:Doob}

We consider the time-homogeneous diffusion process $\left(X^\epsilon_t\right)_{0 \leq t \leq T}$ in $\mathbb{R}^d$ solution of the following stochastic differential equation (SDE)
\be
\label{eq:SDE}
dX^\epsilon_t = F(X^\epsilon_t) \, dt + \sqrt{\epsilon} \, \sigma \circ dW_t \, ,
\ee
where $F \in \mathbb{R}^d$ is a smooth space-dependent drift, $\sigma \in \mathbb{R}^{d \times m}$ is a constant noise matrix and $W_t = \int_0^t dW_s$ is a Brownian motion in $\mathbb{R}^m$ characterised by independent Gaussian increments, i.e., $\mathbb{E}[dW_{t,i}]=0$ and $\mathbb{E}[dW_{t,i} dW_{t,j}] = \delta_{ij} \, dt$ for all $i,j=1,\cdots,m$. Finally, the symbol $\circ$ denotes Stratonovich (mid-point) convention for the stochastic integrals and $\epsilon > 0$ is a parameter, modulating noise intensity, of which we will be mainly interested in situations where $\epsilon \ll 1$. Since $\sigma$ is constant, the It\^{o} and Stratonovich interpretations of the SDE coincide. We nevertheless keep the Stratonovich convention here and in path observables below, where it is necessary to be physically meaningful.

The infinitesimal generator---which acts on $C^2$ bounded functions $\phi$---of the diffusion $\left(X^\epsilon_t\right)_{0 \leq t \leq T}$ is given by
\begin{equation}
    \label{eq:InfGen}
    \mathcal{L}^\epsilon \phi = F \cdot \nabla \phi  + \frac{\epsilon}{2} \nabla \cdot \left( D \nabla \phi  \right) \, ,
\end{equation}
where $D = \sigma \sigma^\top \in \mathbb{R}^{d \times d}$ is a constant diffusion matrix. Formally, the generator locally specifies the transition kernel $P_{0,t}(x,dy) \coloneqq \mathbb{P}(X_t^\epsilon \in dy | X_0^\epsilon = x)$ of the diffusion, viz.\ the probability of moving to $dy$ at time $t$ having started at $x$ at time $0$, as a positive linear operator according to
\begin{equation}
    \label{eq:Semigroup}
    \left( e^{t \mathcal{L}^\epsilon}f \right) (x) =P^\epsilon_{0,t} f(x) \coloneqq \int_{\mathbb{R}^d} P^\epsilon_{0,t}(x,dy) f(y) \coloneqq \mathbb{E} \left[ f(X_t^\epsilon)|X_0^\epsilon=x \right] \, ,
\end{equation}
where the last equality marks a conditional expectation with respect to the transition kernel of the original process.

We set the $\mathbb{R}^d$-valued diffusion in Eq.\ \eqref{eq:SDE} to start at $X^\epsilon_0=x_0$ at time $0$. Additionally, we impose the fixed-time spatial condition $X^\epsilon_T \in \mathcal{R} \subset \mathbb{R}^d$. Note that the initial condition $x_0$ at time $0$ is not special and only simplifies notation. We could be more general and set $X^\epsilon_\tau = x$ at any specific $\tau < T$. By running the diffusion and cutting out trajectories that do not satisfy the final-time constraint, we obtain a specific sub-ensemble of trajectories. The exact same \newtext{configurational path }ensemble can also be obtained by adding to the diffusion in Eq.\ \eqref{eq:SDE} a specific additional drift, which realises the final-conditioned path law dynamically~\cite{Orland2011,Majumdar2015,Corstanje2023}. 

This sets the framework of Doob fixed-time conditioning---the additional drift is written as a function of the scalar positive function $h_t$, also known as Doob's $h$-function, and is defined as
\begin{equation}
    \label{eq:hfunction}
    h_t(x) \coloneqq \mathbb{P}(X_T^\epsilon\in \mathcal{R}|X^\epsilon_t=x) = \mathbb{E} \left[ e^{\epsilon^{-1} k(X_T^\epsilon)} | X_t^\epsilon=x \right] \, ,
\end{equation}
that is the probability for the original process of hitting the target $\mathcal{R}$ at time $T$ conditioned on being at $x$ at time $t$. The second equality in terms of the conditional expectation follows from setting 
\begin{equation}
    \label{eq:BCDoob}
    k(x) =
    \begin{cases}
        0 \hspace{1cm} &\text{if }\; x \in \mathcal{R} \\
        -\infty \hspace{1cm} &\text{otherwise} \\
    \end{cases} \, ,
\end{equation}
guaranteeing the process hits $\mathcal{R}$ at time $T$, and will be useful in the following. The modified diffusion $(\tilde{X}^\epsilon_t)_{0 \leq t \leq T}$ representing the Doob conditioned process is a solution of
\begin{equation}
    \label{eq:DoobSDE}
    d\tilde{X}^\epsilon_t = \left( F(\tilde{X}^\epsilon_t) + \epsilon D \, \nabla \ln h_t(\tilde{X}^\epsilon_t) \right) \, dt + \sqrt{\epsilon} \, \sigma \circ dW_t \, .
\end{equation}

The Doob's process infinitesimal generator $\tilde{\mathcal{L}}^\epsilon$, consequently the transition kernel, can be obtained by Doob transforming the original process infinitesimal generator $\mathcal{L}^\epsilon$ via~\cite{Chetrite2015}
\begin{equation}
    \label{eq:DoobInfGen}
    \mathcal{\tilde{\mathcal{L}}}^\epsilon_t \phi \coloneqq h^{-1}_t \mathcal{L}^\epsilon (h_t \phi) - h^{-1}_t (\mathcal{L}^\epsilon h_t) \phi \, .
\end{equation}
The Doob-transformed process is optimal in the sense that its path measure, generated by $\tilde{\mathcal{L}}^\epsilon_t$ in Eq.\ \eqref{eq:DoobInfGen}, minimises the relative entropy with respect to the path measure of the original process, generated by $\mathcal{L}^\epsilon$ in Eq.\ \eqref{eq:InfGen}, among all stochastic processes realising the conditioning $X_T^\epsilon \in \mathcal{R}$~\cite{Baudoin2002,Lehec2013}. 
\newtext{We remark that this information-theoretic optimality should not in general be identified with minimisation of thermodynamic work or dissipation.}

\deletetext{The Doob $h$-function can be seen as a time-dependent potential, similarly to an external driving, stirring the process to satisfy the final constraint.}\newtext{When the Doob dynamics is physically realised as a controlled process, its additional drift can be interpreted as a time-dependent external control that drives the system towards the final constraint.} Procedurally, one builds this potential as a solution of the backward Kolmogorov equation (BKE)
\begin{equation}
    \label{eq:BackKolmo}
    \left( \partial_t + \mathcal{L}^\epsilon \right) \, h_t = 0 \, ,
\end{equation}
along with the natural boundary condition $h_T(x)=1$ if $x \in \mathcal{R}$ and $0$ otherwise, which enforces hitting the target $\mathcal{R}$ at the final time. Eq.\ \eqref{eq:BackKolmo} follows from the conditional expectation definition in Eq.\ \eqref{eq:hfunction} and by applying the transition kernel operator in Eq.\ \eqref{eq:Semigroup}. The BKE is a linear partial differential equation whose solution is approachable analytically only in few simple cases. We refer the reader to~\cite{Chetrite2015,Majumdar2015} for complementary ways to derive the Doob finite-time conditioned dynamics. 

We remark that, although in the fixed-time conditioning problem considered above the
$h_t$-function is naturally defined as a probability, the framework is not restricted
to conditioning on terminal events. More generally, one may be interested in estimating
exponential moments of the form
\begin{equation}
\label{eq:GenMom}
\mathbb{E}\left[
\exp\left\{-\epsilon^{-1} s K\left[\left(X^\epsilon_t\right)_{0\leq t\leq T}\right]\right\}
\right],
\end{equation}
where $K$ is a suitable time-additive continuous functional of the diffusion path, as in~\cite{Vanden-Eijnden2012}. Additionally, Eq.\ \eqref{eq:GenMom} belongs to the class of Feynman--Kac path measures studied, for example, in~\cite{DelMoral2004}. The exponential reweighting
in Eq.~\eqref{eq:GenMom} defines a \textit{canonical} path ensemble, rather than the
\textit{microcanonical} conditioning described in Eqs.~\eqref{eq:hfunction}
and~\eqref{eq:BCDoob}. In this setting, the relevant $h$-function is no longer a
conditioning probability, but a Feynman--Kac-type exponential moment. Nevertheless,
one can still construct a finite-time Doob-transformed process whose typical
trajectories sample the exponentially tilted distribution. Such trajectories are not
constrained to satisfy a prescribed value of the observable almost surely; instead,
they realise the values selected by the conjugate tilting parameter $s$.

In the following, we report two well-known one-dimensional analytical examples: the Brownian bridge and the Ornstein--Uhlenbeck (OU) bridge, which will be useful for comparison with the results in the last Section of the paper.

\subsubsection*{Brownian bridge}

The classical one-dimensional Brownian bridge is obtained by setting 
$F(x) = 0$ in Eq.\ \eqref{eq:SDE}, then by fixing $X_0=0$ and conditioning the process on reaching a target at time $T$. While we write this formally as $X_T = x_f$, conditioning on a single point is understood in a limiting sense, or equivalently via the Doob's $h$-function in Eq.\ \eqref{eq:hfunction} by identifying the set $\mathcal{R}$ as $dx_f$. In such a case, the $h$-function becomes a transition density and not a probability.

The transition kernel in this example is the usual heat kernel, i.e., $P_{s,t}(x,y) \propto e^{-\frac{(x-y)^2}{2 \epsilon D (t-s)}}$, and the BKE in Eq.\ \eqref{eq:BackKolmo} is explicitly solved by 
\begin{equation}
    \label{eq:hfunctionBrownBridge}
    h_t(x) = \frac{1}{\sqrt{2 \pi \epsilon D (T-t)}} \exp \left( -\frac{(x-x_f)^2}{2 \epsilon D (T-t)} \right) \, .
\end{equation}
Replacing Eq.\ \eqref{eq:hfunctionBrownBridge} into Eq.\ \eqref{eq:DoobSDE} yields
\begin{equation}
    \label{eq:DoobSDEBrownBridge}
    d\tilde{X}^\epsilon_t = -\frac{\tilde{X}^\epsilon_t - x_f}{T-t}  \, dt + \sqrt{\epsilon} \, \sigma dW_t \, .
\end{equation}

\subsubsection*{Ornstein--Uhlenbeck bridge}

The one-dimensional OU bridge is obtained by setting 
$F(x) = -\gamma x$ in Eq.\ \eqref{eq:SDE} and, similarly to before, by fixing $X_0=0$ as well as conditioning the process on reaching $X_T^\epsilon=x_f$. The transition kernel is explicitly given by
\begin{equation}
    P_{0,t}(x,y) = \sqrt{\frac{\gamma}{\pi \epsilon D (1 - e^{-2 \gamma t})}} \exp\left(- \frac{(y-x \, e^{- \gamma t})^2}{\epsilon D(1-e^{-2 \gamma t})}\right)
\end{equation}
and the BKE in Eq.\ \eqref{eq:BackKolmo} is explicitly solved by 
\begin{equation}
    \label{eq:hfunctionOUBridge}
    h_t(x) = \sqrt{\frac{\gamma}{\pi \epsilon D (1-e^{-2 \gamma (T-t)})}} \exp \left( -\frac{(x-x_f \, e^{\gamma (T-t)})^2}{\epsilon D (e^{2 \gamma (T-t)}-1)} \right) \, .
\end{equation}
Replacing Eq.\ \eqref{eq:hfunctionOUBridge} into Eq.\ \eqref{eq:DoobSDE} yields
\begin{equation}
    \label{eq:DoobSDEOUBridge}
    d\tilde{X}^\epsilon_t = - \left( \gamma \tilde{X}^\epsilon_t + 2 \gamma \frac{\tilde{X}^\epsilon_t-x_f \, e^{\gamma (T-t)}}{e^{2 \gamma (T-t)}-1} \right)  \, dt + \sqrt{\epsilon} \, \sigma dW_t \, .
\end{equation}

\begin{figure}[h!]
    \centering
    \includegraphics[width=.8\linewidth]{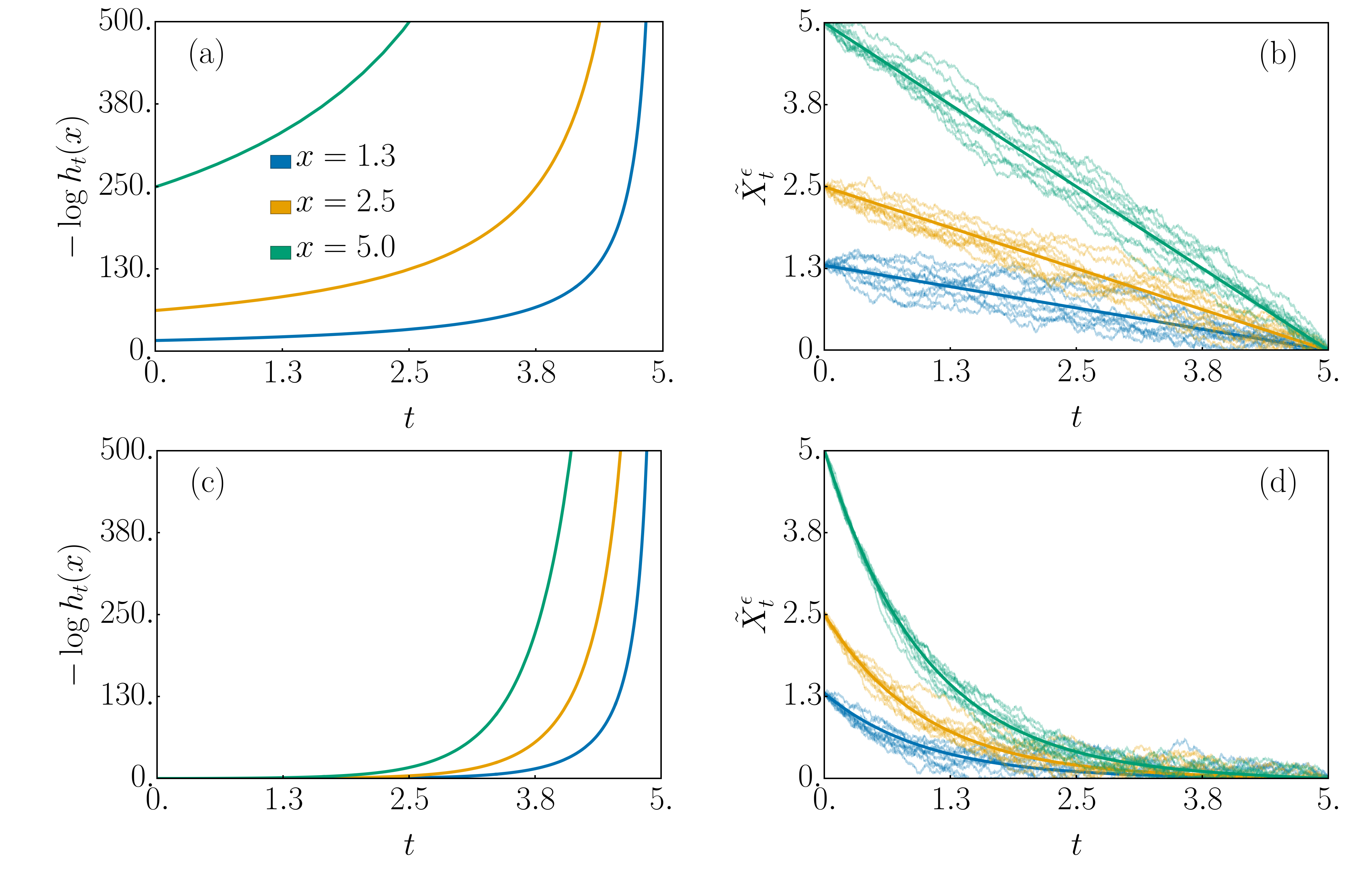}
    \caption{{Panels (a) and (c): rescaled time-dependent potentials $-\log h_t(x)$ (for different initial positions) for Brownian bridge in Eq.\ \eqref{eq:hfunctionBrownBridge} and OU bridge in Eq.\ \eqref{eq:hfunctionOUBridge}. Panels (b) and (d): Doob conditioned trajectories $\tilde{X}_t^{\epsilon}$ representing instances, respectively, of the Brownian Bridge in Eq.\ \eqref{eq:DoobSDEBrownBridge} and Ornstein--Uhlenbeck bridge in Eq.\ \eqref{eq:DoobSDEOUBridge}. Solid lines represent typical realisations $\mathbb{E}[\tilde{X}^{\epsilon}_{t}]$, while shaded areas include real trajectories within one standard deviation. Parameters: $x_f=0,T=5,\epsilon=0.01,\sigma=1.$}}
    \label{fig:Plot_Bridges}
\end{figure}

In Fig.\ \ref{fig:Plot_Bridges} we collect four plots showing $h$-functions and trajectories of Doob conditioned processes. Specifically, the left panels (a) and (c) show rescaled time-dependent potentials $- \ln h_t(x)$ as a function of $t$ with fixed starting value $x$ for, respectively, the Brownian bridge \newtext{in} \deletetext{($h_t(x)$ in }Eq.\ \eqref{eq:hfunctionBrownBridge}\deletetext{)} and OU bridge example \newtext{in} \deletetext{($h_t(x)$ in }Eq.\ \eqref{eq:hfunctionOUBridge}\deletetext{)}. Evidently, the potential sets off at lower values and tend to increase as time goes on, diverging at the final time $T$. It is this specific structure that guarantees the Doob conditioned process to reach target $\mathcal{R}$ at time $T$. Panels (b) and (d) collect stochastic trajectories of these processes, setting off at a specific initial value at time $0$ and targeting $x_f=0$ at time $T$. The typical Brownian bridge trajectory is linear in time, whereas an OU bridge tends to get closer to the bottom of the quadratic potential quickly and stalls there until the final target is reached.

\section{Small-noise large deviations: Hamilton--Jacobi formalism}
\label{sec:Weak}

We focus on path observables of the Doob process with time-additive form
\begin{equation}
    \label{eq:Observable}
    A_{\tau}^T = \int_\tau^T f(\tilde{X}_t^\epsilon) \, dt + \int_\tau^T g(\tilde{X}_t^\epsilon) \circ d\tilde{X}_t^\epsilon \, ,
\end{equation}
which accumulate, linearly with time, state- and current-dependent features of the dynamics~\cite{Jack2010,Chetrite2015,Garrahan2016,Touchette2018}. This class of observables is general and includes physically interesting quantities, such as empirical distributions, empirical currents or flows, work, heat and entropy produced~\cite{Sekimoto2010,Seifert2012}. \newtext{For thermodynamic observables whose definition depends explicitly on the drift or forces driving the process, such as heat and work, their physical interpretation must however be tied to the particular dynamics under consideration. We return to this distinction in Sec.~\ref{subsec:heat}.}

We are interested in studying the whole probability distribution of $A_\tau^T$. We do so by focusing on the moment generating function
\begin{equation}
    \label{eq:MomGenFunDoob}
    Z^{\epsilon}_{\tau,x}(s) = \tilde{\mathbb{E}} \left[ e^{\epsilon^{-1} \, s A_\tau^T} | \tilde{X}_\tau^\epsilon= x \right] \, ,
\end{equation}
where the conditional expectation is taken with respect to the Doob process. This object contains all the information about the typical (zero-noise) behaviour as well as finite-noise fluctuations of the observable $A_\tau^T$ arising by running the Doob process in Eq.\ \eqref{eq:DoobSDE}. Therefore, studying $Z^\epsilon_{\tau,x}$ in Eq.\ \eqref{eq:MomGenFunDoob} requires knowing the Doob's $h$-function solution of the BKE in Eq.\ \eqref{eq:BackKolmo}, which most often is not available analytically. 

Interestingly, $Z^\epsilon_{\tau,x}$ can also be fully derived by focusing on the original process under specific conditions. This more viable path is offered by noticing, once again, that the Doob conditioned process statistically generates the same trajectories that would be obtained by post-selecting trajectories of the original process to satisfy the constraint $X_T^\epsilon \in \mathcal{R}$. Consequently, the conditional expectation with respect to the Doob process in Eq.\ \eqref{eq:MomGenFunDoob} can be replaced by the conditional expectation with respect to the original process, i.e.,
\begin{equation}
    \label{eq:MomGenFunOrigCons}
    Z^\epsilon_{\tau,x}(s) = \mathbb{E} \left[ e^{\epsilon^{-1} \, sA_\tau^T} | X_\tau^\epsilon= x, X_T^\epsilon \in \mathcal{R} \right] = \frac{\mathbb{E} \left[ e^{\epsilon^{-1} \, \left( s A_\tau^T + k(X_T^\epsilon) \right)} | X_\tau^\epsilon= x \right]}{h_\tau(x)} \, ,
\end{equation}
where the second equality follows from conditioning and the second equality in Eq.\ \eqref{eq:hfunction} along with the boundary conditions for $k$ in Eq.\ \eqref{eq:BCDoob}. Notice that, with abuse of notation, $A_\tau^T$ is now \deletetext{interpreted as an observable of the original process}\newtext{evaluated along the post-selected trajectories of the original process while retaining the same path-functional definition. For observables whose physical interpretation depends explicitly on the underlying drift, this representation does not identify the corresponding thermodynamic observable of the Doob-controlled dynamics with that of the original dynamics}. 

For simplicity, we define the unnormalised conditioned moment generating function
\begin{equation}
    \label{eq:UnnormMomGenFunOrigCons}
    \hat{Z}_{\tau,x}^\epsilon(s) \coloneqq \mathbb{E} \left[ e^{\epsilon^{-1} \, \left( s A_\tau^T + k(X_T^\epsilon) \right)} | X_\tau^\epsilon= x \right] \, .
\end{equation}
The second equality in Eq.\  \eqref{eq:MomGenFunOrigCons} can then be rewritten as
\begin{equation}
    \label{eq:MomGenFunOrigConswrtUnnorm}
    Z_{\tau,x}^\epsilon(s) = \frac{\hat{Z}_{\tau,x}^\epsilon(s)}{\hat{Z}_{\tau,x}^\epsilon(0)} \, .
\end{equation}
It is more convenient to work with the unnormalised moment generating function as this allows us to work with the Doob conditioning as a simple terminal penalty/boundary term without the need to carry over the explicit Doob drift.

\newtext{Before proceeding, it is useful to give a path-space interpretation of this reformulation. In the weak-noise regime, trajectories of the original diffusion are exponentially weighted according to their Freidlin--Wentzell action, so that paths departing from the deterministic dynamics carry an exponentially small probability. The factor $e^{\epsilon^{-1}sA_\tau^T}$ reweights this path ensemble according to the value of the observable, while the terminal factor $e^{\epsilon^{-1}k(X^\epsilon_T)}$ imposes the final-time conditioning. The generating function $\hat Z^\epsilon_{\tau,x}(s)$ therefore describes an ensemble of trajectories subject to two simultaneous selections: they must realise the prescribed endpoint event and, depending on $s$, they are biased towards particular fluctuations of $A_\tau^T$. In the weak-noise limit, this reweighted ensemble concentrates on the trajectory, or set of trajectories, that optimally balances the dynamical cost, the observable bias, and the endpoint constraint. The Feynman--Kac and Hamilton--Jacobi construction below provides a systematic way of identifying these dominant paths and their action.}

The unnormalised moment generating function $\hat{Z}^\epsilon_{\tau,x}(s)$ satisfies the following generalised Feynman--Kac relation:
\begin{equation}
    \label{eq:GenFC}
    \left( \partial_\tau + \hat{\mathcal{L}}^\epsilon_s \right) \, \hat{Z}^\epsilon_{\tau,x}(s) = 0 \, ,
\end{equation}
where
\begin{equation}
    \label{eq:TitledGen}
    \hat{\mathcal{L}}^\epsilon_s = F \cdot \left( \nabla + \frac{s}{\epsilon} g \right) + \frac{\epsilon}{2} \left( \nabla + \frac{s}{\epsilon} g \right) \cdot D \left( \nabla + \frac{s}{\epsilon} g \right) + \frac{s}{\epsilon} f \, ,
\end{equation}
is the so-called tilted generator of the process, a linear differential operator generator of the evolution of the unnormalised generating function of $A_\tau^T$. Additionally, the Feynman--Kac equation is to be solved backward in time starting from the terminal boundary condition $\hat{Z}_{T,x}^\epsilon(s)=e^{\epsilon^{-1} k(x)}$. The Feynman--Kac relation for $g=0$ (no jump contributions) is a classic result in probability theory~\cite{Kac1949,Fitzsimmons1999,RogersWilliams2000} and the extended version including jump contributions was also considered more recently in the large deviation literature~\cite{Lebowitz1999,Bertini2015,Chetrite2015}. Its form, along with the tilted generator, can be obtained in multiple ways. We refer the reader to~\cite{Chetrite2015} for a derivation making use of Girsanov theorem. In Appendix~\ref{app:multi}, we propose a more pedestrian derivation by using It\^{o}'s calculus only. \newtext{Furthermore, we remark that equations of the type \eqref{eq:GenFC} provide the starting point of the logarithmic-transform approach to stochastic control originally developed in~\cite{Fleming1982}, which is similar in spirit to what we do in the following.}

By fixing a value for the tilting parameter $s$, the tilted operator $\hat{\mathcal{L}}^\epsilon_s$ biases the path measure at the level of the moment generating function towards trajectories conditioned to satisfy, on average, a specific value for the observable $A_\tau^T$~\cite{Jack2010,Chetrite2015,Garrahan2016,Touchette2018}. Na\"{i}vely, for $s=0$, $A_\tau^T$ will concentrate exponentially with $\epsilon^{-1}$ around the typical value. For $s \neq 0$, $A_\tau^T$ will concentrate on atypical values and the specific duality relation between $s$ and $A_\tau^T$ is given below. \newtext{[}We remark that if we focused on the normalised version of the generating function, i.e., $Z_{\tau,x}^\epsilon$, then the drift $F$ in the tilted generator in Eq.\ \eqref{eq:TitledGen} would need to be replaced by $F + \epsilon D \nabla \ln h_t$ and the boundary term for Eq.\ \eqref{eq:GenFC} would become $Z_{T,x}^\epsilon = 1$.\newtext{]}

Similarly to~\cite{Fleming1982,Fleming1985,Vanden-Eijnden2012}, by replacing $\hat{Z}_{\tau,x}^\epsilon(s)$ with its log (Hopf--Cole) transform, i.e.,
\begin{equation}
    \label{eq:HopfColeTransf}
    \hat{M}_{\tau,x}^\epsilon(s) = \epsilon \log \hat{Z}_{\tau,x}^\epsilon(s) \, , 
\end{equation}
in the generalised Feynman--Kac relation \eqref{eq:GenFC}, after some algebra, we obtain the following second-order Hamilton--Jacobi (HJ) equation:
\begin{equation}
    \label{eq:SecHJ}
    \partial_\tau \hat{M}_{\tau,x}^\epsilon(s) = - H(x, \nabla \hat{M}_{\tau,x}^\epsilon(s) + s g(x)) - \frac{\epsilon}{2} \nabla \cdot \left[ D (\nabla \hat{M}^\epsilon_{\tau,x} + sg(x)) \right]  \, ,
\end{equation}
with terminal condition $\hat{M}_{T,x}^\epsilon(s) = k(x)$ where
\begin{equation}
    \label{eq:HamHJ}
    H(x,p+sg) = F(x) \cdot (p + sg) + \frac{1}{2} p \cdot D p + s p \cdot D g + \frac{s^2}{2} g \cdot D g + sf \, .
\end{equation}

In the weak-noise limit $\epsilon \rightarrow 0$, the second-order HJ equation reduces to the first-order HJ equation of the form
\begin{equation}
    \label{eq:FirHJ}
    \partial_\tau \hat{M}_{\tau,x}(s) = - H(x, \nabla \hat{M}_{\tau,x}(s) + s g) \, ,
\end{equation}
where
\begin{equation}
    \label{eq:Action}
    \hat{M}_{\tau,x}(s) = \lim_{\epsilon \rightarrow 0 } \hat{M}_{\tau,x}^\epsilon(s) \, ,
\end{equation}
is the so-called zero-viscosity approximation of $\hat{M}^\epsilon_{\tau,x}$ (see~\cite{Fleming2006} for details). We call this object action as it is the leading exponential behaviour of the unnormalised moment generating function, i.e., 
\begin{equation}
    \hat{Z}_{\tau,x}^\epsilon(s) \asymp e^{\epsilon^{-1} \hat{M}_{\tau,x}(s)} \, ,   
\end{equation}
where $\asymp$ marks equality at logarithmic scale. In the statistical physics literature, the approximation in the display above is also known as WKBJ (after Wentzel, Kramers, Brillouin
and Jeffreys) ansatz~\cite{Assaf2017,Weber2017}. By returning to normalised objects using $M_{\tau,x}^\epsilon(s) = \hat{M}_{\tau,x}^\epsilon(s) - \epsilon \ln h_\tau(x)$, the finite-time weak-noise scaled cumulant generating function (SCGF) takes the form
\begin{equation}
    \label{eq:SCGF}
    M_{\tau,x}(s) = \hat{M}_{\tau,x}(s) - \hat{M}_{\tau,x}(0) \, ,
\end{equation}
where we generally assume the form 
\begin{equation}
	\label{eq:ConnhM}
	h_\tau(x) \asymp e^{\epsilon^{-1} \hat{M}_{\tau,x}(0)} \, .
\end{equation}

Generally speaking, the solution of the first-order HJ equation can be obtained from the characteristic curves of the associated Hamiltonian dynamics subject to appropriate boundary conditions. By integrating along these trajectories, say $(x_t^*,p_t^*)_{\tau \leq t \leq T}$, the action in Eq.\ \eqref{eq:Action} can be rewritten in the form
\begin{equation}
    \label{eq:ActionChar}
    \hat{M}_{\tau,x}(s) = k(x^*_T) - \int_\tau^T \left[ p_t^* \cdot \dot{x}_t^* - H(x_t^*,p_t^* + s g(x^*_t)) \right] \, dt \, ,  
\end{equation}
with $x^*_\tau \equiv x$ and $x^*_T \equiv x_f$.

Solving Eq.\ \eqref{eq:FirHJ} for $\hat{M}_{\tau,x}$ via its characteristics is often analytically impractical for complex systems and one needs to resort to numerical methods (see, e.g.,~\cite{Grafke2019}). Here instead, we turn to a Lagrangian perspective. By applying a Legendre transform to $H(x,p+sg)$, we obtain the Lagrangian density 
\begin{align}
\nonumber
    L(x_t,\dot{x}_t) &= \frac{1}{2} \norm{\dot{x}_t - F(x_t) - s D  g(x_t)}^2_D   - s F(x_t) \cdot g(x_t) - \frac{s^2}{2} g(x_t) \cdot D g(x_t) - s f(x_t) \\
    \label{eq:Lagrangian}
    &= \frac{1}{2} \norm{\dot{x}_t - F(x_t)}^2_D   - s \dot{x}_t \cdot g(x_t) - s f(x_t)
    \, ,
\end{align}
where $\norm{a}^2_b \coloneqq a \cdot b^{-1} a$, which, when replaced into the action \eqref{eq:ActionChar}, yields the variational form
\deletetext{\begin{equation}
    \label{eq:ActionOptimal}
    \hat{M}_{\tau,x}(s) = \sup_{\substack{x_t \in \mathcal{AC}([\tau,T]) \\ x_\tau = x}} \left( k(x_T) - \int_\tau^T \left(  L(x_t,\dot{x}_t)  \right) \, dt \right) \, ,
\end{equation}}
\newtext{\begin{equation}
    \label{eq:ActionOptimal}
    \hat{M}_{\tau,x}(s) = \sup_{\substack{x_t \in \mathcal{AC}([\tau,T]) \\ x_\tau = x}} \left( k(x_T) - \int_\tau^TL(x_t,\dot{x}_t)\, dt \right) \, ,
\end{equation}}
which has to be maximised for trajectories $x_t$ that are absolutely continuous in the time interval $[\tau,T]$. \newtext{Eq.\ \eqref{eq:ActionOptimal} makes the path-space picture introduced above explicit. Its optimiser $x_t^*$ is the most probable trajectory, at exponential order in the weak-noise limit, among trajectories satisfying the endpoint conditioning and biased towards the fluctuation of $A_\tau^T$ selected by $s$. For $s=0$ it gives the dominant trajectory of the conditioned ensemble, whereas $s \neq 0$ selects atypical realisations of the observable within that same ensemble. The resulting instantons therefore describe the dynamical mechanism by which a fluctuation occurs.}

The action in Eq.\ \eqref{eq:ActionOptimal} represents the unnormalised SCGF for an observable of the form \eqref{eq:Observable} in the weak-noise limit. Its normalised form, obtained by a simple shift, is given in Eq.\ \eqref{eq:SCGF}. From the SCGF, we can derive the function $I_{\tau,x}(a)$ via the Legendre--Fenchel transform
\begin{equation}
    \label{eq:RateFunction}
    I_{\tau,x}(a) = \sup_{s \in \mathbb{R}} \left( s a - M_{\tau,x}(s) \right) \, ,
\end{equation}
which via the G\"{a}rtner--Ellis theorem~\cite{DenHollander2000,Touchette2009,Dembo2010} gives the (convex envelope of) the rate function of the large deviation principle for the Doob conditioned distribution
\begin{equation}
    \label{eq:LDP}
    \mathbb{P}(A_\tau^T \in da | X_\tau^\epsilon = x, X_T^\epsilon = x_f) \asymp e^{-\epsilon^{-1} I_{\tau,x}(a)} \, .
\end{equation}

Finally, when the variational problem has a unique optimiser and $\hat{M}_{\tau,x}$ is differentiable, the optimiser associated with $s$ gives the typical path $x^*_t$ realising the corresponding biased fluctuation of the Doob process $A_\tau^T = a$ linked to $s=s^*$ via the Legendre duality
\begin{equation}
    \label{eq:Duality}
    \partial_s \hat{M}_{\tau,x}(s)|_{s=s^*} = a \, .
\end{equation}
\newtext{Multiple competing optimisers would correspond to distinct mechanisms realising the same conditioned fluctuation and may signal coexistence between dynamical pathways.} \deletetext{Evidently}\newtext{Finally}, large deviation functions and probabilities are directly influenced by the normalisation $\hat{M}_{\tau,x}(0)$, connected to the Doob's $h_t$-function in Eq.\ \eqref{eq:ConnhM}, however, the instanton paths are not, since for fixed $(\tau,x)$ the term $\hat{M}_{\tau,x}(0)$ is simply a constant. 

\subsection{Heat dissipated by \newtext{the} Doob\newtext{-controlled}\deletetext{a}  \deletetext{conditioned process}\newtext{dynamics}}\label{sec:dissipated_Heat}
\label{subsec:heat}

We now specialise the general framework to the heat dissipated by a \deletetext{Doob-conditioned}\newtext{Doob-controlled} process. Heat provides a natural bridge between stochastic dynamics and thermodynamics: at the trajectory level, it coincides, up to a factor of the bath temperature, with the entropy change of the environment, while its fluctuations encode irreversibility through the log-ratio of forward and time-reversed path probabilities~\cite{Sekimoto2010,Seifert2012}. Work is another natural observable whose fluctuations have recently been studied using Feynman--Kac theory~\cite{Mamede_2024}, although not in the context of Doob-conditioned dynamics. Here, we use heat as a first observable to examine the thermodynamics of Doob\deletetext{conditioning}\newtext{-controlled dynamics}. Our aim is not to exhaust the thermodynamic content of Doob-\deletetext{conditioned}\newtext{controlled} processes, but to initiate its study with the methods developed above, leaving a systematic treatment to future work.

\newtext{A distinction is necessary between the heat exchanged by the original dynamics along trajectories post-selected on the terminal condition and the heat dissipated by the Doob dynamics when its additional drift is interpreted as an external control force~\cite{Jack2016}. Although post-selection and Doob control generate the same conditioned ensemble of trajectories for the configurational degrees of freedom, thermodynamic observables that depend explicitly on the force driving the dynamics need not coincide. In particular, the additional Doob drift performs work on the system, and its contribution therefore enters the heat dissipated by the controlled dynamics.}

\newtext{Here we specialise to this latter quantity.} The observable \eqref{eq:Observable} becomes the heat dissipated, equivalently the medium entropy contribution, for a Doob\newtext{-controlled} \deletetext{conditioned} process for $f=0$ and 
\begin{equation}
    \label{eq:gDoobCond}
    g(x) = 2 D^{-1} (F(x) + \epsilon D \nabla \ln h_t(x)) \, ,
\end{equation}
where the Doob's $h_t$-function is defined in Eq.\ \eqref{eq:hfunction} and we allow $g$ to explicitly depend on time but suppress this dependence in the notation. Using \eqref{eq:ConnhM} in the weak-noise limit, Eq.\ \eqref{eq:gDoobCond} can be written as
\begin{equation}
    \label{eq:gDoobCondWeak}
    g(x) = 2D^{-1} (F(x) + D \nabla \hat{M}_{t,x}(0)) \, ,
\end{equation}
where $\hat{M}_{t,x}(0)$ is a solution of the maximisation problem in Eq.\ \eqref{eq:ActionOptimal} for $s=0$ and for all $t \in [\tau,T]$.

For an absolutely continuous path of the Doob\newtext{-controlled} dynamics and using $g$ in \eqref{eq:gDoobCondWeak}, we can then define the dissipated heat as
\begin{equation}
    \label{eq:DissipatedHeatWeakNoise}
    Q_\tau^T = 2 \int_\tau^T (F(x_t) + D \nabla \hat{M}_{t,x_t}(0)) \cdot D^{-1} \dot{x}_t \, dt \, .
\end{equation}
\newtext{Accordingly, $Q\tau^T$ should be understood as the heat dissipated by the controlled dynamics realising the Doob-conditioned ensemble, rather than the heat of the original dynamics evaluated along post-selected trajectories.} Fluctuations of this object can be studied by solving Eq.\ \eqref{eq:ActionOptimal} with the specific Lagrangian
\begin{equation}
    \label{eq:LagrangianHeat}
    L(x_t,\dot{x}_t) = \frac{1}{2} \norm{\dot{x}_t - F(x_t)}^2_D - 2 s (F(x_t)  + D \nabla \hat{M}_{t,x_t}(0)) \cdot D^{-1} \dot{x}_t \, .
\end{equation}
Then, the SCGF in Eq.\ \eqref{eq:SCGF} is connected to the rate function $I_{\tau,x}(q)$ via the Legendre--Fenchel transform \eqref{eq:RateFunction} for $a \equiv q$. 

\section{Illustrative examples}\label{sec:ex}

\subsection{Area observable for an one-dimensional Brownian Bridge}\label{sec:Area_Observable_Example}
As a first illustrative example, we consider a one-dimensional Brownian bridge, whose dynamics, connecting $X^\epsilon_0=x$ with $X^\epsilon_T=x_f$, is given by Eq.~\eqref{eq:DoobSDEBrownBridge} by setting $D=\sigma=1$. For such a process, we study small noise large deviations for the linear time-integrated observable
\begin{equation}
\label{eq:Observable_Linear_BB}
A_0^T = \int_0^T X_t^\epsilon\ dt \, ,
\end{equation}
obtained from the general observable in Eq.\ \eqref{eq:Observable} by fixing $\tau=0$, $f(x) = x$, and $g(x)=0$. Such a Brownian functional can be geometrically interpreted as the signed area accumulated between the stochastic bridge path and the time axis, in the spirit of related Brownian-area functionals studied for constrained Brownian paths~\cite{Majumdar2005}.

The Lagrangian~\eqref{eq:Lagrangian} containing information on small-noise fluctuations regardless of the specific Doob conditioning takes the following explicit form:
\begin{equation}
    \label{eq:Lagrangian_First_Example}
    L(x_t,\dot{x}_t) = \frac{1}{2} \dot{x}_t^2 - s x_t \, .
\end{equation}
Consequently, the action takes the form  
\begin{equation}
    \label{eq:Action_First_Example}
    \hat{M}_{\tau,x}(s) = \sup_{\substack{x_t \in \mathcal{AC}([0,T]) \\ x_0 = x \\ x_T = x_f}}  \int_0^T \left( {sx_t} - \frac{1}{2} \dot{x}_t^2 \right) \, dt \, .
\end{equation}
We remark that the optimiser for $s=0$ represents the typical trajectory of the Doob conditioned process whereas for $s>0$ and $s<0$ the optimal trajectory is such to favor, respectively, positive and negative fluctuations of the total area. Finally, the SCGF is obtained from Eq.\ \eqref{eq:SCGF}. Such an object contains all the information about fluctuations of the Brownian bridge: its first derivative identifies typical values of accumulated area for specific biases, the second derivative around $s=0$ is the asymptotic (in noise) variance, and higher derivatives tell about higher moments.

\begin{figure}[htb!]
    \centering
    \includegraphics[width=1\linewidth]{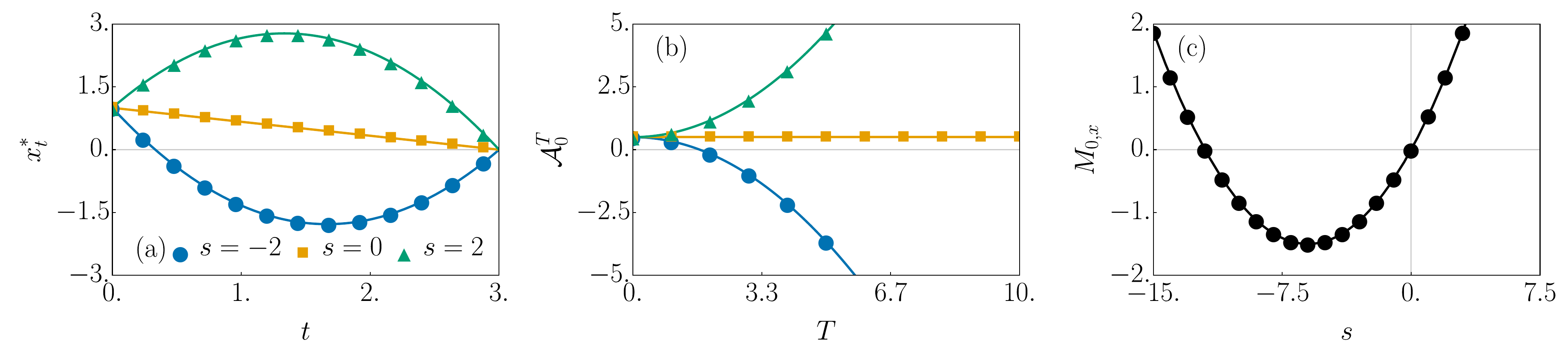}
    \caption{Panel (a): Analytical optimal trajectories \eqref{eq:optimal_sol_area} (solid lines) compared with numerical solutions (markers) of Eq.\ \eqref{eq:Action_First_Example} for exemplary values of the tilting parameter $s$. Panel (b): Analytical form of the area rate observable \eqref{eq:CumArea_Final_Obs} (solid lines) compared with numerical optimisation (markers) of Eq.\ \eqref{eq:Action_First_Example} then replaced in \eqref{eq:Observable_Linear_BB} for the same exemplary values of $s$ as in panel (a). Panel (c): analytical SCGF $M_{0,x}(s)$ from Eq.\ \eqref{eq:SCGF_Area} (solid line) and numerical SCGF from Eq.\ \eqref{eq:Action_First_Example}, shifted by $\hat{M}_{0,x}(0)$ (markers). Parameters used: $x=1$, $x_f=0$, and $D=1$; $T=3$ for panels (a) and (c).}
    \label{fig:Area_Observable}
\end{figure}

The Euler–Lagrange equation associated to Eq.\ \eqref{eq:Action_First_Example} is
\begin{equation}
\ddot{x}_t + s = 0 \, ,
\end{equation}
which is to be solved subject to the bridge boundary conditions $x_0 = x$ and $x_T = x_f$. Solving this boundary value problem yields the optimal trajectory
\begin{equation}
\label{eq:optimal_sol_area}
x^{*}_t = \frac{s}{2}t(T - t) + \frac{t (x_f - x)}{T} + x \, ,
\end{equation}
reported in Fig.\ \ref{fig:Area_Observable}(a) for three exemplary values of $s$ as well as for $x=1$, $x_f=0$, and $T=3$. 

Replacing the optimal trajectory directly in the observable \eqref{eq:Observable_Linear_BB} and renormalising by the total time $T$ gives
\begin{equation}
\label{eq:CumArea_Final_Obs}
    \mathcal{A}_0^T \coloneqq \frac{A_0^T}{T}=\frac{s}{12}T^2+\frac{x+x_f}{2} \, ,
\end{equation}
which is the mean rate at which area accumulates in a Brownian bridge process and is plotted in Fig.\ \ref{fig:Area_Observable}(b). Similarly, using Eq.\ \eqref{eq:Action_First_Example} along with Eq.\ \eqref{eq:optimal_sol_area}, the action takes the exact form
\begin{equation}
	\label{eq:Shift_Area}
    \hat{M}_{0,x}(s)=\frac{s^2T^3}{24}+\frac{sT}{2}(x+x_f)-\frac{(x_f-x)^2}{2T},
\end{equation}
and, consequently, the SCGF in Eq.\ \eqref{eq:SCGF} is
\begin{equation}
\label{eq:SCGF_Area}
    M_{0,x}(s)=\frac{s^2 T^3}{24}+\frac{sT}{2}(x+x_f) \, .
\end{equation}
This quantity is then plotted in Fig.\ \ref{fig:Area_Observable}(c).

By inspecting Fig.\ \ref{fig:Area_Observable}(b), in the typical scenario $s=0$ the Brownian bridge has only the endpoint triangular contribution to the area (cfr.\ Fig.\ \ref{fig:Plot_Bridges}(b)). In the fluctuation regime instead, the area rate $\mathcal{A}_0^T$ in Eq.\ \eqref{eq:CumArea_Final_Obs} scales quadratically with time. The Legendre duality \eqref{eq:Duality} can be verified directly by deriving the SCGF \eqref{eq:SCGF_Area} by $s$, which gives
\begin{equation}
    \label{eq:LegDualityArea}
    a(s) = \frac{sT^3}{12} + \frac{T}{2}(x+x_f) \, ,
\end{equation}
connecting the tilting parameter $s$ with the typical fluctuation value for the observable $A_0^T$. For $s=0$ and after renormalising by the total time $T$, we obtain the constant contribution from Eq.\ \eqref{eq:CumArea_Final_Obs}. Atypical values of $\mathcal{A}_0^T$ can also be investigated by inverting the relation \eqref{eq:LegDualityArea}.

\subsection{Heat dissipated \newtext{by a Doob-controlled}\deletetext{in an} Ornstein--Uhlenbeck process}

We consider the one-dimensional OU process with $F(x) = -\gamma x$, where \( \gamma>0 \) is the relaxation rate, and $D = \sigma^2=1$ in the SDE \eqref{eq:SDE}. The OU process is conditioned to visit $X^\epsilon_T = x_f$.

We focus on the total heat dissipated \newtext{by the corresponding Doob-controlled dynamics}, which takes the stochastic form in Eq.\ \eqref{eq:Observable} with $f=0$ and $g$, in the weak-noise limit, from Eq.\  \eqref{eq:gDoobCondWeak}. Heat fluctuations are characterised by first solving the variational problem in Eq.\ \eqref{eq:ActionOptimal} with the Lagrangian in Eq.\ \eqref{eq:LagrangianHeat}. The Euler--Lagrange equation is
\begin{equation}
    \label{eq:Gen_EL_2D}
    \ddot{x}_t-\gamma^2 x_t = 2s \partial_t \partial_{x} \hat{M}_{t,x_t}(0) \, .
\end{equation}

In the typical case \(s=0\), the optimal path connecting \(x\) to \(x_f\) is explicitly given by
\begin{equation}
    x^*_t
    =
    \frac{\sinh[\gamma(T-t)]}{\sinh (\gamma T)}
  x
    +
\frac{\sinh (\gamma t)}{\sinh (\gamma T)}
      x_f \, .
\end{equation}
From this solution, the corresponding Doob correction is given by
\begin{equation}
\label{eq:General_Doob_2D}
    \partial_{x} \hat M_{t,x_t}(0)
    =
    \gamma
    \left[
    \bigl(1-\coth[\gamma(T-t)]\bigr) x_t
    +
    \operatorname{csch}[\gamma(T-t)]
     x_f
    \right] \, .
\end{equation}

Replacing Eq.\  \eqref{eq:General_Doob_2D} into Eq.\ \eqref{eq:Gen_EL_2D} and taking the time derivative on the r.h.s., we obtain the general Euler--Lagrange equation
\begin{equation}
    \label{eq:EL_GeneralOU}
    \ddot{x}_t-\gamma^2[1-2s\csch^2[\gamma(T-t)]] x_t
    =
\frac{2\gamma^2s\coth[\gamma(T-t)]}{\sinh[\gamma(T-t)]}   x_f .
\end{equation}
The \deletetext{general} \newtext{finite-action} solution \newtext{obtained by setting the bridge boundary conditions $x^*_T=x_f$ and $x_0^*=x_0$} is given by 
\begin{equation}
    \label{eq:InstantonHeat1D}
    x^*_{t}
    =
    x_{f} \cosh [\gamma(T-t)]
    +
    \left[
    x_{0}-x_{f}\cosh[\gamma T]
    \right]
    \frac{
    Q_{\nu_s}^1(\coth [\gamma(T-t)])
    }{
    Q_{\nu_s}^1(\coth [\gamma T])
    } \, ,
\end{equation}
where $Q_{\nu}^{\mu}(z)$ denotes the associated Legendre function of the second kind~\cite{Olver2010}, with degree \(\nu\), order \(\mu\), and the standard real branch for \(z>1\). It is a solution of
\begin{equation}
\label{eq:AssLegEq}
(1-z^2)y''-2zy'
+\left[\nu(\nu+1)-\frac{\mu^2}{1-z^2}\right]y=0 \, .
\end{equation}
In our case $\mu=1$ and
\begin{equation}
    \nu_s=-\frac{1}{2}+\frac{1}{2}\sqrt{1-8s} \, ,
\end{equation}
which, in turns, set the finite-action real branch to exist only for $s<1/8$ (some discussions on this follow). We collect in Appendix~\ref{sec:Derivation_NonConservative_Appendix} the full derivation of the instanton equation \eqref{eq:InstantonHeat1D}\newtext{, along with the details on the regularity conditions at the terminal time together with the correct selection of the physical branch.}

\begin{figure}[htb!]
    \centering
    \includegraphics[width=1\linewidth]{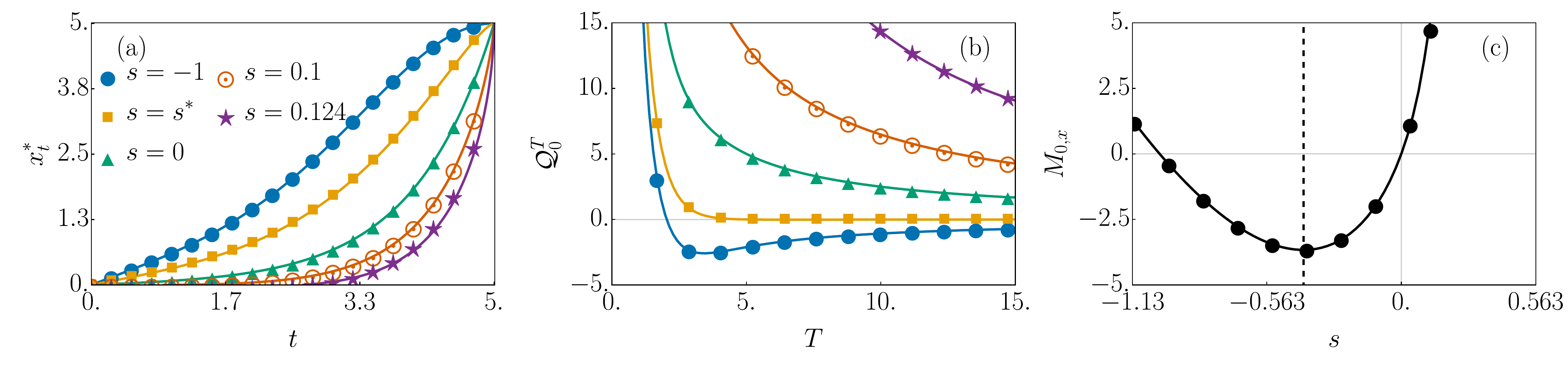}
    \caption{Panel (a): Analytical optimal trajectories \eqref{eq:InstantonHeat1D} (solid lines) compared with numerical solutions (markers) of the variational problem for exemplary values of the tilting parameter $s$. Panel (b): Semi-analytical form for the average rate of heat dissipation \newtext{under the Doob-controlled dynamics} from Eq.\ \eqref{eq:Final_Heat_1D_MT} (solid lines) compared with numerical solutions (markers) for the same exemplary values of $s$ as in panel (a). Panel (c): Semi-analytical SCGF from Eq.\ \eqref{eq:SCGF} compared with numerical solutions (markers). The dashed line marks the value $s=s_c$, minimum of the SCGF. Parameters used: \deletetext{$x=0$}\newtext{$x_0=0$}, $x_f=5$, and $D=1$; $T=5$ for panels (a) and (c). }
    \label{fig:OU1DHeat}
\end{figure}

Instantons are plotted in Fig.\ \ref{fig:OU1DHeat}(a) for several values of the tilted parameter $s$.  The mean rate of dissipated heat \newtext{by the Doob-controlled dynamics along}\deletetext{over} these instanton trajectories can be calculated using \deletetext{the instanton }\eqref{eq:InstantonHeat1D} and the definition of heat in Eq.\ \eqref{eq:DissipatedHeatWeakNoise}. Eventually, it takes the form
\begin{equation}
\label{eq:Final_Heat_1D_MT}
    \mathcal{Q}_0^T \coloneqq \frac{Q_0^T}{T} = \frac{2\gamma}{T}\int_{0}^{T}\left[ \csch{[\gamma(T-t)]x_{f}-\coth{[\gamma(T-t)]}x_{t}^*}\right] \dot{x}_{t}^*\ dt \, ,
\end{equation}
and representative curves are plotted in Fig.\ \ref{fig:OU1DHeat}(b). Evidently, the rare trajectories that connect the bottom of the potential with a value far away from it in a fixed time $T$ will typically dissipate heat---the larger the value of $s$, the greater the heat dissipated---under the Doob protocol. These instanton trajectories tend to favor more heat dissipation by staying closer to the bottom of the potential for longer and delaying the excursion to reach the target as can be seen from Fig.\ \ref{fig:OU1DHeat}(a). Conversely, atypical trajectories with negative $s$ tend to reduce the heat dissipated by staying far away from the bottom of the OU potential for longer. For $s < s_c$, a critical value that we identify in the following, such a behaviour is supported by absorbing heat from the environment for values of the conditioning-time $T$ larger than a certain threshold (which depends on the target value $x_f$). 

This behaviour is evidently very different from a standard OU process. \newtext{This contrast arises because the former includes the additional Doob control force, whereas the latter does not.} Normally, trajectories of an OU process that climb up the quadratic potential necessarily absorb heat from the environment---namely, build up fluctuations thanks to the white noise that move the process away from the bottom of the potential. Otherwise, no climbing can happen. In fact, the only way for such an OU process to dissipate heat would be the opposite behaviour, viz.\ slide down the potential reaching for the bottom. The behaviour of a\deletetext{n} \newtext{Doob-controlled} OU \deletetext{bridge}\newtext{process} is very different because the Doob's $h$-function acts as time-dependent drift changing the properties of the overall effective potential acting on the system. As we see here, heat is typically dissipated in climbing the time-independent background potential and the \deletetext{OU bridge can be made to}\newtext{Doob drift may even} act as a refrigerator by pushing the process to stay far away from the bottom of the potential for the longest. 

Eventually, the action associated with the instanton solution \eqref{eq:InstantonHeat1D} is written as follows:
\begin{equation}
    \label{eq:ActionHeat1D}
    \hat{M}_{0,x}(s)=-\int_{0}^{T}\left[\frac{1}{2}\left( \dot{x}^*_{t}+\gamma x_{t}^* \right)^2-2\gamma s\dot{x}_{t}^* \left(\csch{[\gamma(T-t)]x_{f}-\coth{[\gamma(T-t)]}x_{t}^*}\right)\right] \, d t \, .
\end{equation}
Using the last display along with Eq.\ \eqref{eq:General_Doob_2D}, the SCGF in \eqref{eq:SCGF} can also be calculated and we plot its form in Fig.\ \ref{fig:OU1DHeat}(c). As $s \rightarrow 1/8$, the SCGF diverges, implying that
the associated rate function---obtained via the Legendre--Fenchel transform in Eq.\ \eqref{eq:RateFunction}---is asymptotically linear with slope $1/8$ for positive fluctuations of heat dissipated (cfr.\ Fig.\ 5~\cite{Touchette2009}). Additionally, we identify a threshold value $s=s_c<0$, which is implicitly a function of $T$ and $x_f$, that marks a sign change in the heat dissipated: from positive for $s>s_c$ to negative for $s < s_c$. 

As noticed, the parameter space is large. The OU constants $\gamma$ and $D$, the starting value $x$, the final value $x_f$, the final conditioning-time $T$, and the tilting parameter $s$ all play a role in characterising the dynamics. However, it is beyond the scope of this work to understand the details of this specific model and we set these aspects aside for future work.

\subsection{Folding of a Biomolecule}

To \newtext{further} illustrate \deletetext{other physical}\newtext{the thermodynamic} implications of the Doob-\deletetext{conditioned}\newtext{controlled} dynamics, we consider a minimal model of biomolecular folding described by two reaction coordinates. These variables can represent collective degrees of freedom associated with the conformational state of the molecule along the folding pathway~\cite{Onuchic1997EnergyLandscape}. We model the folding landscape by a two-dimensional potential with two metastable minima, associated with unfolded and folded conformations, separated by a saddle point corresponding to the transition state. The potential is given by
\begin{equation}
    u(x_1,x_2)=(x_1^2-1)^2+x_2^2+2x_1 x_2\deletetext{.}\newtext{ ,}
\end{equation}
\newtext{where we will mark as $x_A$ and $x_B$ its minima. More specifically, $x_A=(-\sqrt{3/2},\sqrt{3/2})$ and $x_B=(\sqrt{3/2},-\sqrt{3/2})$.}

We consider trajectories conditioned to start at $X^\epsilon_t=x_A$ at $t=0$ and to reach $X^\epsilon_T=x_B$ at a prescribed final time $T$. This bridge condition \deletetext{can be interpreted as an experimental protocol in which the molecule is required to complete the folding transition within a finite observation time, thereby selecting trajectories that are typically rare under the original dynamics.}\newtext{selects trajectories in which the molecule completes the folding transition within a prescribed finite time, trajectories that are typically rare under the original dynamics. If the conditioned ensemble is realised dynamically through the Doob transform, the additional drift can instead be interpreted as a control protocol driving the molecule towards the target conformation.}

The overdamped dynamics of the original process is governed by Eq.\ \eqref{eq:SDE} with $X^{\epsilon}_t=(X^{\epsilon}_{1,t},X^{\epsilon}_{2,t})^\top\in\mathbb{R}^2$ the state vector, $F = - \nabla u $, $W_t=(W_{1,t},W_{2,t})^\top$ a standard two-dimensional Brownian motion with independent components. We further assume an isotropic diffusion matrix $D=\sigma\sigma^\top= \text{I}$. Although simplified, this landscape captures the essential ingredients of activated folding dynamics: metastability and barrier crossing.

We focus once again on the dissipated heat in Eq.\ \eqref{eq:DissipatedHeatWeakNoise},  which quantifies the thermodynamic cost associated with \newtext{the control protocol} enforcing a (rapid) folding transition. The Lagrangian \eqref{eq:LagrangianHeat} is explicitly given by 
\begin{equation}
\label{eq:Folding_Lagrangian}
L(x_{t},\dot{x}_{t})=\frac{1}{2}\norm{\dot{x}_t-F(x_t)}^2_{\text{I}}
-2s\dot{x}_t\cdot(F(x_t)+\nabla\hat{M}_{t,x_t}(0)),
\end{equation}
with $x_t\equiv(x_{1,t},x_{2,t})^T$, leading to the following expression for $\hat{M}_{0,x}(s)$ in the following form
\begin{equation}
\label{eq:Biomolecule}
    \hat{M}_{0,x_A}(s) = \sup_{\substack{x_t \in \mathcal{AC}([\tau,T]) \\ x_\tau = x_A \\ x_T=x_B}} \left( - \int_{\tau}^T\left(\frac{1}{2} \norm{\dot{x}_t -F(x_t)}^2_{\text{I}} - 2s \dot{x}_t \cdot (F(x_t) + D \nabla \hat{M}_{t,x_t}(0))\right)\,dt  \right) \, .
\end{equation}

\begin{figure}[htb!]
    \centering
    \includegraphics[width=1.\linewidth]{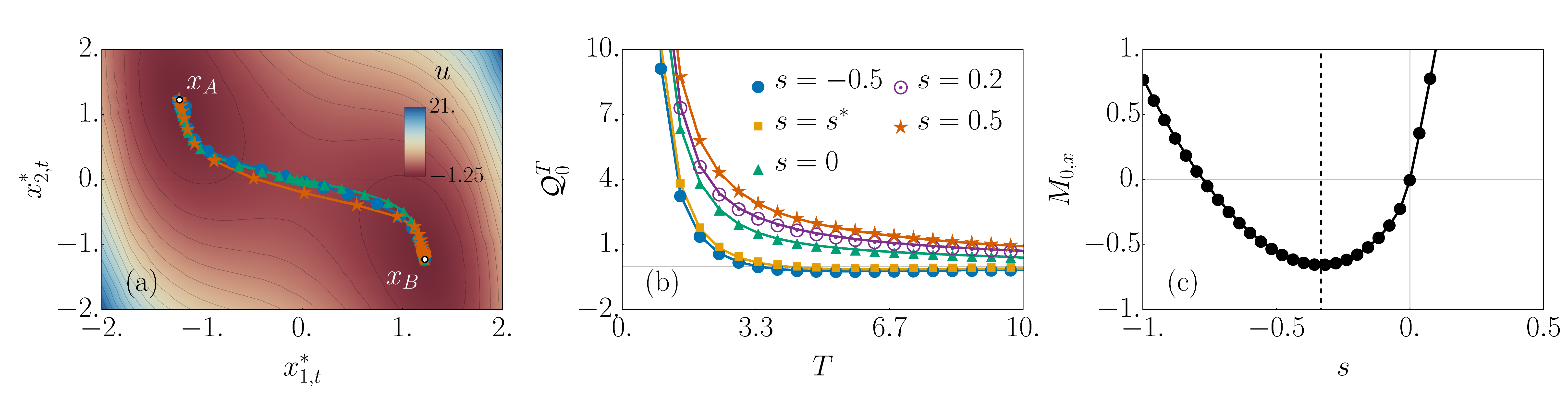}
    \caption{Panel (a): Numerical solutions of instantons obtained solving the variational problem in Eq.\ \eqref{eq:Biomolecule} for exemplary values of the tilting parameter $s$. Panel (b): Numerical values for the mean rate of \deletetext{the }heat dissipation\newtext{ under the Doob-controlled dynamics} from Eq.\ \eqref{eq:DissipatedHeatWeakNoise} using numerical instantons from \eqref{eq:Biomolecule}. Panel (c): Numerical SCGF from Eqs.\ \eqref{eq:SCGF} and \eqref{eq:Biomolecule}. The dashed line marks the minimum of the SCGF $s=s^*$. Parameters used: \deletetext{$x_A=(-\sqrt{3/2},\sqrt{3/2})$, $x_B=(\sqrt{3/2},-\sqrt{3/2})$, and} $D=1$\deletetext{;} \newtext{and, specifically,}  $T=4$ for panels (a) and (c)}
    \label{fig:Biomol_Scheme}
\end{figure}

The optimal trajectories describing the most probable transition paths connecting the two metastable states for several values of $s$ are plotted in Fig.\ \ref{fig:Biomol_Scheme}(a) and the mean rate of dissipated heat \newtext{under the Doob control} is plotted in Fig.\ \ref{fig:Biomol_Scheme}(b) as a function of the final time $T$. For the parameters considered here and for $s \leq 0$, viz.\ for typical and smaller than typical values of heat dissipated, instantons cross the saddle point of the background potential $u$, whereas higher dissipation is obtained staying away from it. \newtext{These trajectories therefore identify how the folding mechanism changes upon selecting different heat fluctuations.}

\newtext{The instanton trajectories are obtained by direct numerical optimisation of the variational problem in Eq.\ \eqref{eq:Biomolecule} following the optimal-control formulation in Ref.~\cite{Vanden-Eijnden2012}. We introduce an uniform temporal grid with $t_i=i\Delta t$, $i=1,\dots N$ and $\Delta t=T/N$ and discretise trajectories as $\{x_i\}_{i=0}^{N}$. Both endpoints $x_0=x_A$ and $x_{N}=x_B$ are kept fixed and the internal values are treated as optimisation variables. The continuous action is replaced by its correspondent finite-dimensional discrete version and, therefore, minimised. The family of solutions as a function of the bias parameter $s$ is computed by continuation. Starting from the optimal trajectory at $s=0$, the converged solution at $s=s_k$ was used as an initial guess for the optimisation for a next value $s=s_{k+1}$. Both branches of $s>0$ and $s<0$ are calculated separately starting at $s=0$.}

\section{Conclusion}
\label{sec:conc}

In this work we studied rare events of finite-time Doob conditioned diffusion processes in the weak-noise regime. Starting from the observation that the Doob conditioned path ensemble coincides with the ensemble of trajectories of the original process post-selected on the terminal constraint, we reformulated the moment generating function of time-additive observables without explicitly relying on the Doob drift. This allowed us to encode the conditioning through a terminal boundary term and to derive a generalised Feynman--Kac equation for the unnormalised generating function. In the weak-noise limit, this equation reduces to a Hamilton--Jacobi problem whose solution gives the leading exponential contribution to the generating function.

We then showed that the corresponding action admits a Lagrangian variational representation, turning the problem of characterising fluctuations of conditioned processes into an optimal-control problem \newtext{in the variational sense}. The normalised finite-time weak-noise SCGF is obtained by subtracting the contribution at zero bias, which is connected to the leading exponential behaviour of the Doob's $h$-function. We illustrated the framework on analytically tractable bridge processes and on heat dissipation in Doob\newtext{-controlled} \deletetext{conditioned} dynamics. For the Brownian bridge, the area observable provides a simple benchmark for the variational formalism. For the Ornstein--Uhlenbeck bridge, the heat observable leads to a non-trivial instanton equation whose solutions show how the Doob protocol reshapes dissipation along rare conditioned trajectories. Finally, we applied the same construction to a minimal biomolecular folding model, showing how the method can be used numerically in higher-dimensional systems where analytical expressions for the Doob drift are not available.

There are several natural directions for future work. A first one is to extend the formalism to genuinely time-dependent drifts, where the notation becomes more involved but the same post-selection principle should still apply. A second direction is to study many-particle systems, where the inverse-noise large deviation scaling is replaced by the number of particles. Interestingly, in Ref.~\cite{Grela2021}, the Doob $h$-function was calculated analytically for a system of $N$ vicious Brownian bridges by mapping the problem to non-intersecting Dyson Brownian bridges. It would be interesting to apply our approach to this setting and investigate fluctuations of time-additive observables within such conditioned many-particle ensembles. More generally, the variational representation developed here suggests direct optimisation schemes for conditioned fluctuations in systems where the $h$-function is inaccessible, including high-dimensional models and interacting diffusions. Finally, it would be interesting to connect this weak-noise bridge formalism more closely with thermodynamic questions, \deletetext{in order to understand not only which rare trajectories are selected by a terminal constraint, but also the energetic and entropic cost of producing them}\newtext{in particular by systematically comparing thermodynamic observables of the original dynamics along post-selected trajectories with those of the controlled Doob dynamics that realises the same configurational path ensemble}.

\section*{Acknowledgments}
IM acknowledges financial support from the Brazilian agency FAPESP under grant 2023/17704-2 and from the Special Research Fund (BOF) of Hasselt University under Grant No. BOF25BL12. FC is supported by a Leverhulme Early Career Fellowship No.\ ECF-2025-482. The authors are grateful to Juan P.\ Garrahan and Kay Brandner for insightful discussions.

\appendix

\section{Derivation of the generalised Feynman--Kac relation via It\^{o}'s calculus}
\label{app:multi}

In this Section we show in a pedagogical way how to derive the Feynman--Kac Eqs.\ \eqref{eq:GenFC}--\eqref{eq:TitledGen} for time-additive observables that include both state and current dependent terms. We keep the derivation general here and focus on multiplicative noise, only eventually restricting to additive noise to show the connection with the main text. The Feynman--Kac equation for only state-dependent observables is a classical result in probability theory~\cite{Kac1949,Fitzsimmons1999,RogersWilliams2000} whereas the inclusion of current-like contributions is more recent and discussed in~\cite{Lebowitz1999,Bertini2015,Chetrite2015}. Although these equations are known in the literature, we offer here a detailed pedagogical derivation.

We consider the Stratonovich SDE with multiplicative noise that follows:
\begin{equation}
    \label{eq:SDEMultiplicative}
    dX_t^\epsilon = F(X_t^\epsilon)dt + \sqrt{\epsilon}  \, \sigma (X_t^\epsilon) \circ dW_t \, ,
\end{equation}
where, similarly to Eq.\ \eqref{eq:SDE}, $F \in \mathbb{R}^d$ is a smooth space-dependent drift, $\sigma \in \mathbb{R}^{d \times m}$ is a smooth space-dependent noise matrix and $W_t$ is a Brownian motion in $\mathbb{R}^m$ characterised by independent Gaussian increments. We denote by $D = \sigma \sigma^\top$ the space-dependent diffusion matrix.

The first step is to transform the Stratonovich SDE into an It\^{o} SDE. This is readily done using the standard conversion
\begin{equation}
    \label{eq:StratoToIto}
    \sigma(X_t^\epsilon) \circ dW_t 
    = 
    \sigma(X_t^\epsilon) dW_t 
    + 
    \frac{1}{2} \sum_{j=1}^m d [\sigma_{\cdot j}(X^\epsilon_t),W_j]_t \, ,
\end{equation}
where the last term on the r.h.s.\ is the so-called quadratic covariation, contracted over the Brownian components. Generally, given two $\mathbb{R}^d$-valued processes $X_t$ and $Y_t$, such that
\begin{equation}
	\label{eq:SDEXY}
	dX_t = a_t dt + b_t dW_t, 
	\qquad 
	dY_t = c_t dt + d_t dW_t \, ,
\end{equation}
the quadratic covariation is defined infinitesimally as
\begin{equation}
\label{eq:QuadraticCovariation}
d[X,Y]_t = dX_t dY_t^\top \, .
\end{equation}
Therefore, using \eqref{eq:SDEXY} we obtain
\begin{equation}
\label{eq:QuadraticCovariationFinal}
d[X,Y]_t = b_t d_t^\top dt \, ,
\end{equation}
as all other terms are $o(dt)$. To apply this to Eq.\ \eqref{eq:StratoToIto}, we need to replace component-wise $X_t \rightarrow \sigma_{\cdot j}(X_t^\epsilon)$ and $Y_t \rightarrow W_j$, and then sum over $j=1,\dots,m$. By It\^{o}'s lemma we write $d\sigma(X_t^\epsilon)$ and extract the Brownian part, which is
\begin{equation}
\sqrt{\epsilon} \, \nabla \sigma(X_t^\epsilon) \sigma(X_t^\epsilon) dW_t \, .
\end{equation}
Hence, we get
\begin{equation}
\label{eq:StratoToItoQuadratic}
\sum_{j=1}^m d [\sigma_{\cdot j}(X^\epsilon_t),W_j]_t
=
\sqrt{\epsilon} \,
\left[
\nabla \cdot D
-
\sigma \nabla \cdot \sigma^\top
\right](X_t^\epsilon) dt \, .
\end{equation}
The Stratonovich SDE can therefore be written in the It\^{o} form
\begin{equation}
    \label{eq:ItoSDEPre}
    dX_t^\epsilon 
    =
    \left[
    F
    +
    \frac{\epsilon}{2}
    \left(
    \nabla \cdot D
    -
    \sigma \nabla \cdot \sigma^\top
    \right)
    \right](X_t^\epsilon)dt
    +
    \sqrt{\epsilon}\sigma(X_t^\epsilon)dW_t \, .
\end{equation}
Equivalently, defining the corrected drift
\begin{equation}
    \label{eq:ModifiedDrift}
    \hat{F}(x) 
    =
    F(x)
    -
    \frac{\epsilon}{2}
    \left(
    \sigma(x)\nabla\cdot\sigma^\top(x)
    \right) \, ,
\end{equation}
we obtain the It\^{o} SDE
\begin{equation}
    \label{eq:ItoSDE}
    dX_t^\epsilon 
    =
    \hat{F}(X_t^\epsilon) dt 
    + 
    \frac{\epsilon}{2} 
    (\nabla \cdot D)(X_t^\epsilon) dt 
    + 
    \sqrt{\epsilon} \sigma(X_t^\epsilon) dW_t \, . 
\end{equation}
The associated infinitesimal generator is then written as
\begin{equation}
    \label{eq:InfGenMult}
    \mathcal{L}^\epsilon f(x) = \hat{F}(x) \cdot \nabla f(x) + \frac{\epsilon}{2} \nabla \cdot \left( D(x) \nabla f(x) \right) \, .
\end{equation}

The next step is to write the It\^{o} differential for the observable $A_t^u$ in Eq.\ \eqref{eq:Observable}, which, combined with the equation above, gives
\begin{align}
    \nonumber
    dA_t^u &= f(X_u^\epsilon) du + g(X_u^\epsilon) \cdot dX_u^\epsilon + \frac{1}{2} d [g(X_u^\epsilon),X_u^\epsilon] \\
    &= (f +  g \cdot \hat{F} + \frac{\epsilon}{2} g \cdot (\nabla \cdot D) )(X_u^\epsilon) du + \sqrt{\epsilon} g(X_u^\epsilon) \cdot \sigma(X_u^\epsilon) dW_u  + \frac{1}{2} d [g(X_u^\epsilon),X_u^\epsilon] \\
    &= (f +  g \cdot \hat{F} + \frac{\epsilon}{2} g \cdot (\nabla \cdot D) + \frac{\epsilon}{2} \text{Tr} [D \nabla g]  )(X_u^\epsilon) du + \sqrt{\epsilon} \, g(X_u^\epsilon) \cdot \sigma(X_u^\epsilon) dW_u \\
    &= (f +  g \cdot \hat{F} + \frac{\epsilon}{2} \nabla \cdot (D g)) (X_u^\epsilon) du + \sqrt{\epsilon} \, g(X_u^\epsilon) \cdot \sigma(X_u^\epsilon) dW_u \, ,
        \label{eq:ItoObs}
\end{align}
where the second-to-last step follows from It\^{o}'s lemma for $g(X_u^\epsilon)$, viz.\ 
\begin{align}
    \label{eq:ItosLemmag}
    dg(X_t^\epsilon) &= \left( \partial_t g + \mathcal{L}^\epsilon g \right)(X_t^\epsilon)  dt + \sqrt{\epsilon} \, (\nabla g)^\top \sigma(X_t^\epsilon) dW_t 
\end{align}
and from using the quadratic covariation formula in Eq.\ \eqref{eq:QuadraticCovariationFinal}. 

The next step is to calculate the differential of $\hat{Z}^\epsilon_{u,X_t^\epsilon}(s)$ defined in Eq.\ \eqref{eq:UnnormMomGenFunOrigCons}, which will be useful in the following. This is readily done by using It\^{o}'s lemma and takes the form
\begin{equation}
    \label{eq:DiffMom}
    d \hat{Z}^\epsilon_{u,X_t^\epsilon} = (\partial_u \hat{Z}^\epsilon_{u,X_u^\epsilon} + \mathcal{L}^\epsilon \hat{Z}^\epsilon_{u,X_u^\epsilon}) du + \sqrt{\epsilon} \nabla (\hat{Z}^\epsilon_{u,X_u^\epsilon})^\top \sigma(X_u^\epsilon) dW_u \, . 
\end{equation}
This describes the stochastic evolution of the future value function $\hat{Z}^\epsilon_{u,X_u^\epsilon}$ along trajectories of the original process. However, $\hat Z^\epsilon_{u,x}$ contains only the exponential weight accumulated from $u$ to the final time. When $u$ is advanced to $u+du$, the object $\hat{Z}^\epsilon_{u+du,X_{u+du}^\epsilon}$ contains the remaining weight from $u+du$ onward, and therefore misses the infinitesimal contribution accumulated on the interval $[u,u+du]$. Thus the path-weighted object is not $\hat Z^\epsilon_{u,X_u^\epsilon}$ alone, but rather its exponentially reweighted version, in which this short-time contribution is inserted explicitly. Consequently, applying Itô's lemma to $\hat Z^\epsilon_{u,X_u^\epsilon}$ alone only produces the original generator $\mathcal L^\epsilon$. The Feynman--Kac tilted operator arises only after applying Itô's lemma to the exponentially weighted product, which keeps track of both the weight already accumulated and the remaining future weight.

To do so, for $u \geq t$, we use the additivity of the observable, i.e., $A_t^T = A_t^u + A_u^T$, and introduce the object
\begin{align}
    \label{eq:YDefApp}
    Y_u &\coloneqq \mathbb{E} \left[ e^{\epsilon^{-1} \, \left( s A_t^T + k(X_T^\epsilon) \right)} | \mathcal{F}_u \right] \\
    &= \exp (\epsilon^{-1} s A_t^u ) \hat{Z}^\epsilon_{u,x} \, ,
\end{align}
where $\mathcal{F}_u$ is the whole information up to time $u$. The display above shows the explicit exponential reweighting of the (unnormalised) moment generating function. Its differential is
\begin{equation}
    \label{eq:DiffYu}
    dY_u = \exp (\epsilon^{-1} s A_t^u ) d \hat{Z}^\epsilon_{u,x} + d \left[ \exp (\epsilon^{-1} s A_t^u ) \right]  \hat{Z}^\epsilon_{u,x} + d \left[ \exp (\epsilon^{-1} s A_t^u ),  \hat{Z}^\epsilon_{u,x}  \right] \, .
\end{equation}

By It\^{o}'s lemma, we write
\begin{align}
    d \left[ \exp (\epsilon^{-1} s A_t^u ) \right] &= \exp (\epsilon^{-1} s A_t^u ) \left( \epsilon^{-1} s \, d A_t^u + \frac{1}{2} \epsilon^{-2} s^2 \, d [A_t^u,A_t^u] \right) \\
    \label{eq:ItoExp}
    \begin{split}
   &\hspace{-2.25cm}= \exp (\epsilon^{-1} s A_t^u ) \bigg( \epsilon^{-1} s \left[ (f +  g \cdot \hat{F} + \frac{\epsilon}{2} \nabla \cdot (D g) )(X_u^\epsilon) du + \sqrt{\epsilon} g(X_u^\epsilon) \cdot \sigma(X_u^\epsilon) dW_u \right] + \\
    &\hspace{2cm}+ \frac{1}{2} \epsilon^{-1} s^2 g(X_u^\epsilon)^\top D(X_u^\epsilon) g(X_u^\epsilon) du \bigg) 
    \end{split}
\end{align}
where the second equality follows by substituting Eqs.\ \eqref{eq:ItoObs} and 
\begin{equation}
     d[A_t^u,A_t^u] = \epsilon \,g(X_u^\epsilon)^\top D(X_u^\epsilon) g(X_u^\epsilon) du \, .
\end{equation}

Substituting Eqs.\ \eqref{eq:DiffMom} and \eqref{eq:ItoExp} into Eq.\ \eqref{eq:DiffYu} along with the quadratic covariation
\begin{align}
    d \left[ \exp (\epsilon^{-1} s A_t^u ),  \hat{Z}^\epsilon_{u,x}  \right] &= s \exp (\epsilon^{-1} s A_t^u ) g(X_u^\epsilon)^\top D(X_u^\epsilon) \nabla (\hat{Z}^\epsilon_{u,X_u^\epsilon}) \, du \, ,
\end{align}
gives the explicit form of $dY_u$. Now, by using the fact that 
\begin{equation}
    Y_u = \mathbb{E} [Y_{u + du} |X_u^\epsilon] \, ,
\end{equation}
we have that
\begin{equation}
    \mathbb{E} \left[ dY_u |X_u^\epsilon \right] = 0 \, ,
\end{equation}
viz.\ $Y_u$ is a martingale. Taking the conditional expectation of Eq.\ \eqref{eq:DiffYu}, the $dW_u$ terms vanish and after factoring out the common positive factor $\exp (\epsilon^{-1} s A_t^u )$ we are left with
\begin{equation}
    \begin{split}
    \partial_u \hat{Z}^\epsilon_{u,X_u^\epsilon} &+ \mathcal{L}^\epsilon \hat{Z}^\epsilon_{u,X_u^\epsilon} + \epsilon^{-1} s \bigg[ (f +  g \cdot \hat{F} + \frac{\epsilon}{2} \nabla \cdot (D g) )(X_u^\epsilon) + \\
    &\frac{1}{2}  s g(X_u^\epsilon)^\top D(X_u^\epsilon) g(X_u^\epsilon) \bigg] \hat{Z}^\epsilon_{u,X_u^\epsilon} + s g(X_u^\epsilon)^\top D(X_u^\epsilon) \nabla (\hat{Z}^\epsilon_{u,X_u^\epsilon}) = 0 \, ,
    \end{split}
\end{equation}
which recovers the Feynman--Kac equation in \eqref{eq:GenFC} and \eqref{eq:TitledGen} in the additive noise case setting $\sigma$ and $D$ constant and, therefore, $\hat{F}=F$.

\section{Instanton calculation for the Ornstein--Uhlenbeck example} \label{sec:Derivation_NonConservative_Appendix}

We start from Eq.\ \eqref{eq:EL_GeneralOU}, which for the sake of readability is reported here as well
\begin{equation}
    \ddot{x}_{\tau}-\gamma^2[1-2s\csch^2[\gamma(T-\tau)]]x_{\tau}=\frac{2\gamma^2 sx_{f}\coth[\gamma(T-\tau)]}{\sinh[\gamma(T-\tau)]} \, .
    \label{eq:eq_POSCHTELLER_Appendix_1D}
\end{equation}
\newtext{We seek for a solution subject to the bridge boundary conditions $x^\star_0=x_0$ and $x^\star_T=x_f$.}

Eq.\ \eqref{eq:eq_POSCHTELLER_Appendix_1D} is in the form of a hyperbolic Pöschl–Teller equation with a source term (the r.h.s.)~\cite{Olver2010}. To solve it, we start by making the following change of \deletetext{variable}\newtext{variables} $u \coloneqq \gamma(T-\tau)$ \newtext{and $y_u \coloneqq x_{T-u/\gamma}$}, which leads to
\newtext{\begin{equation}
    \ddot{y}_{u}-[1-2s\csch^2{u}]y_{u}=\frac{2 sx_{f}\coth{u}}{\sinh{u}} \, ,
\label{eq:eq_POSCHTELLER_Appendix_1D_sub_u}
\end{equation}}
\deletetext{\begin{equation}
    \ddot{x}_{u}-[1-2s\csch^2{u}]x_{u}=\frac{2 sx_{f}\coth{u}}{\sinh{u}} \, .
\label{eq:eq_POSCHTELLER_Appendix_1D_sub_u}
\end{equation}}
\newtext{with boundary conditions $y_{\gamma T}=x_0$ and $y_0=x_f$.}

\deletetext{To}\newtext{We} further simplify the equation\deletetext{, we carry out a second change of variables} \newtext{by considering} $z \coloneqq \coth{u}$ \newtext{and, as a consequence,  $w_z \coloneqq y_{\operatorname{arcoth}z}$}. Derivatives will change as follows:
\begin{equation}
\label{eq:z_Relations_Appendix}
    \dv{}{u}=-(z^2-1)\dv{}{z},\quad \dv[2]{}{u}=(z^2-1)^2\dv[2]{}{z}+2z(z^2-1)\dv{}{z},\quad \csch^2{u}=z^2-1 \, ,
\end{equation}
and \eqref{eq:eq_POSCHTELLER_Appendix_1D_sub_u} further simplifies into
\deletetext{\begin{equation}
\label{eq:Posch_Teller_z_Appendix}
    (1-z^2)\ddot{w}_{z}-2z\dot{w}_{z}-\left[2s+\frac{1}{1-z^2}\right]w_{z}=\mathcal{F}_z \, .
\end{equation}}
\newtext{\begin{equation}
\label{eq:Posch_Teller_z_Appendix}
    (1-z^2)\ddot{w}_{z}-2z\dot{w}_{z}-\left[2s+\frac{1}{1-z^2}\right]w_{z}= \frac{2 s x_f z}{\sqrt{z^2 - 1}} \, .
\end{equation}}
\deletetext{with the source term given by \deletetext{$\mathcal{F}_z \coloneqq 2 s x_{\coth T} \, z / \sqrt{z^2 - 1}$}\newtext{$\mathcal{F}_z \coloneqq 2 s x_f \, z / \sqrt{z^2 - 1}$}.} \newtext{The transformed domain and boundary conditions become
\begin{equation}
    z\in\left[\coth(\gamma T),\infty
    \right),\qquad \begin{cases}
        w_{\coth{\gamma T}}=x_0,\\
        \lim_{z\to\infty}w_z=x_f \, .
    \end{cases}
\end{equation}
}

The solution of the homogeneous part is obtained by comparison with the associated Legendre equation in \eqref{eq:AssLegEq}, 
leading to
\deletetext{\begin{equation}
    x^{(h)}_{z}=A_1P_{\nu_s}^{1}(z)+A_2Q_{\nu_s}^{1}(z),
\end{equation}}
\newtext{\begin{equation}
    w^{(h)}_{z}=A_1P_{\nu_s}^{1}(z)+A_2Q_{\nu_s}^{1}(z),
\end{equation}}
where apex $(h)$ stands for `homogeneous', \deletetext{$P$}\newtext{$P^{1}_{\nu_s}$} and \deletetext{$Q$}\newtext{$Q^{1}_{\nu_s}$} are the associated Legendre functions of the first and second kind \newtext{with order $\mu=1$ and degree $\nu_s$}, respectively, $A_1$ and $A_2$ constants to be fixed by boundary conditions, and 
\begin{equation}
    \label{eq:nu_s}
    \nu_s=-\frac{1}{2}+\frac{\sqrt{1-8s}}{2} \, .
\end{equation}
Then, the source term allows for the particular (apex $(p)$) solution
\deletetext{\begin{equation}
    x^{(p)}_{z}=x_{\coth T}\frac{z}{\sqrt{z^2 - 1}} \, ,
\end{equation}}
\newtext{\begin{equation}
    w^{(p)}_{z}=x_{f}\frac{z}{\sqrt{z^2 - 1}} \, ,
\end{equation}}
which can be verified straightforwardly. \newtext{Thus,
\begin{equation}
    \omega_z
    =
    A_1P_{\nu_s}^{1}(z)
    +
    A_2Q_{\nu_s}^{1}(z)
    +
    x_f\frac{z}{\sqrt{z^2-1}}.
\end{equation}}

\deletetext{Finally, the general}\newtext{The final }solution is simply given by the sum of both homogeneous and particular solutions, yielding in the original time frame
\begin{equation}
\label{eq:Solution_1DOU_Appendix}
    x^{*}_{\tau}=A_1P_{\nu_s}^{1}(\coth[\gamma(T-\tau)])+A_2Q_{\nu_s}^{1}(\coth[\gamma(T-\tau)])+x_{f}\cosh[\gamma(T-\tau)] \, ,
\end{equation}
with a slight abuse of notation for the constants.

Setting the endpoint bridge condition $x^*_{T} = x_{f}$ immediately implies $A_1=0$ to avoid divergences as $\tau \rightarrow T$. Then, imposing the boundary at the starting point, i.e., $x^*_{0} = x_{0}$, the second constant is also fixed as
\begin{equation}
    \label{eq:A2}
    A_2 = \frac{x_{0}-x_{f} \cosh [\gamma T]}{Q_{\nu_s}^1(\coth [\gamma T])} \, ,
\end{equation}
which substituting back into \eqref{eq:Solution_1DOU_Appendix} yields the final instanton Eq.\ \eqref{eq:InstantonHeat1D}.

To conclude, we carry out a further check to make sure that the associated Legendre function of the second kind does not lead to any divergencies in the finite-time action $\hat{M}_{0,x}(s)$ in Eq.\ \eqref{eq:ActionHeat1D}. As $\tau \to T$, 
\begin{equation}
    Q_{\nu_s}^1(\coth [\gamma (T-\tau)])\sim (T-\tau)^{\nu_s+1} \, .
\end{equation}

The contribution to the action of Eq.\ \eqref{eq:ActionHeat1D} is therefore dominated by 
\begin{equation}
    \hat{M}_{0,x}(s)
    \propto
    \int_0^{T} (T-t)^{2 \nu_s}\,dt \, .
\end{equation}
This integral is finite if and only if $\nu_s> -1/2$, which implies $s < 1/8$, well within the existence condition for Eq.\ \eqref{eq:nu_s}. At the critical value $s=1/8$, the action diverges and also for $s > 1/8$, when the instanton develops complex behaviour.

\section*{References}
\bibliography{Bib052023.bib}

\end{document}